\documentclass[smallextended]{svjour3}                    
\smartqed  
\usepackage{graphicx,subfig,nicefrac,multirow,natbib}
\begin{document}

\title{Sequential Monitoring of a Bernoulli Sequence when the Pre-change Parameter is Unknown}

 \author{Gordon J. Ross \and
Dimitris K. Tasoulis \and
Niall M. Adams}

\institute{Gordon J. Ross \at
              Department of Mathematics\\
              Imperial College London\\
              London, SW7 2PG\\
              Tel.: +44(0)2075940990\\
              Fax: +44(0)2075940923\\
              \email{gordon.ross03@imperial.ac.uk}           
}


\date{Received: date / Accepted: date}

\maketitle

\begin{abstract}

The task of monitoring for a change in the mean of a sequence of Bernoulli random variables has
been
widely studied. However most existing approaches make at least one of the
following assumptions, which may be violated in many real-world situations: 1) the
pre-change value of the Bernoulli parameter is known in advance, 2) computational
efficiency is not paramount, and 3) enough observations occur between
change points to allow asymptotic approximations to be used. We develop a novel change detection method based on Fisher's Exact Test which does not make
any of these assumptions. We show that our method can be implemented in a computationally
efficient manner, and is hence suited to sequential monitoring where new
observations are constantly being received over time. We assess our method's performance empirically via using simulated data, and find that it is comparable to the
optimal CUSUM scheme which assumes both pre- and post-change values of the
parameter to be known.
\end{abstract}

\section{Introduction}
We are concerned with the task of detecting changes in the mean of an equally spaced, discrete-time sequence of Bernoulli($\theta_t$) random variables. In most traditional work on this probem, the observed data is assumed to be a fixed length sequence of realizations $x_1,\ldots,x_t$ from the independent random variables $X_1,\ldots,X_t$. Here, $t$ is a known integer and is usually interpreted as denoting the time at which the observation was made. Recently, much work has instead focused on the case where $t$ is not fixed \citep{Woodall1997}, and the observations instead constitute a \textbf{data stream} \citep{Domingos2003} of an unknown and potentially infinite length. This data stream setting arises when new observations are being constantly received over time, for example in high frequency finance \citep{Chen1997}, or more traditional industrial monitoring problems where the task is to continually monitor the output of a manufacturing process to detect an increase in the number of defective items \citep{Yeh2008}. When working with data streams it is not known in advance how many observations will be received, and the fixed length assumption is hence inappropriate. We will refer to this data stream setting as \textbf{sequential monitoring}, and the remainder of the paper focuses on this case.  As a point of terminology, we will refer to the $i^{th}$ observation
as being a \textit{success} if $X_i=1$, and a \textit{failure} if $X_i = 0$

We assume that the value of the Bernoulli parameter $\theta_t$ is constant but unknown between each change point, and that the distribution of each of the $X_i$ variables can hence be written as:

$$
X_t \sim \textrm{Bernoulli}(\theta_t), \quad \theta_t = \left\{ \begin{array}{rl}
\theta_0, &\mbox{ if $t \leq \tau_1$} \\
\theta_1, &\mbox{ if $\tau_1 < t \leq \tau_2$}\\
\theta_2, &\mbox{ if $\tau_2 < t \leq \tau_3$}\\
...
       \end{array} \right.
$$
where each value of $\tau_i$ denotes a change point. The task is to detect
any changes which take
place, as soon after they occur as possible, and to accurately
estimate the location of each change point. In sequential change detection, it is standard to try and detect each change point in sequence. 
 A change detector is run until the first change is encountered. After this has been found, the detector is restarted and monitoring for the second change point begins. In this way, the task is reduced to the successive detection of individual change points, and the problems associated with attempting to detect multiple change points simultaneously are avoided. Therefore for the remainder of the paper we will treat the stream as if it contained at most a single change point, with the understanding that our techniques could also be deployed on streams containing several.

The performance of sequential change detectors is usually measured using two
criteria \citep{Basseville1993}: the expected time between false positive
detections (denoted
$ARL_0$, for Average
Run Length), and the
mean delay until a change is detected (denoted $ARL_1$). If $\tau$ denotes the true location of the change point, and $\hat{\tau}$ is the time at which the change detector signalled for a change, then we can define these formally as:

$$ARL_0 = E(\hat{\tau}|\theta=\theta_0), \quad ARL_1 = E(\hat{\tau}-\tau|\theta=\theta_1).$$
When $\hat{\tau}<\tau$, a false positive is said to have occurred. These quantities play a similar role to the Type I and Type II errors in traditional hypothesis testing. It is considered very important \citep{Basseville1993} for a sequential change detector to have a known bound on the $ARL_0$ so that change points can be assessed for statistical significance, for the same reason that hypothesis tests are required to have a bound on the Type I error probability.


Within the traditional statistics literature, the analysis of the Bernoulli change detection problem has mainly focused on the case where the sequence $X_1,\ldots,X_t$ has a fixed length \citep{Pettitt1980,Hinkley1970}. These approaches suffer from several limitations which makes them inappropriate tools for the sequential monitoring problem. First, they can be computationally inefficient when used in such a context, since they require all previous data to be stored in memory, with test
statistics being recomputed from scratch whenever a new observation is
received. Second, although these approaches do provide a way to bound the Type I error when testing for a change in a fixed length sequence, this does not easily translate into a method for bounding the $ARL_0$, which is required in sequential monitoring problems. Finally, they often use test statistics which only asymptotically have
known distributions. Using asymptotic theory is valid when it can
be assumed that there are many observations between each change point. However
when changes are occuring frequently, the asymptotic distribution may diverge
substantially from the exact small-sample distribution.

Much work on the non-fixed length problem 
comes from the Quality Control literature. When the Bernoulli random variables can be naturally split up into groups of size $n>1$ and hence treated as Binomial random variables, the \textit{p}-chart is
the standard tool used to monitor for change \citep{Montgomery2005}. However, the
\textit{p}-chart is not ideal when the observations are arriving individually and $n=1$, a situation commonly referred to as continuous inspection.

Existing change detection methods for the $n=1$ case can be categorized into those which are transformation-based, and those which use the untransformed observations. An example of the 
former is \cite{Nelson1994}, which treats the time between failures as defining a sequence of Geometric random variables, and then uses a transformation to make this approximately Gaussian. Standard tools for detecting a change in the parameters of a Gaussian sequence can then be deployed.. Several approaches have been proposed which do not rely on transformations.
Versions of the popular CUSUM chart which can be deployed on untransformed
observations are proposed in
\cite{Reynolds1999} and \cite{Chang2001}. A method using EWMA
charts is also considered in \cite{Yeh2008}.

A key limitation of these existing approaches is that they assume the
pre-change value
of the Bernoulli parameter, $\theta_0$, is known exactly. 
However this is an idealization that may not be applicable in several real world
situations, including those mentioned above. Although in some situations it may be possible to estimate $\theta_0$ from a reference sample which is known to come from the pre-change distribution, this will not always be the case. The performance of change detection algorithms under misspecification of the pre-change distribution has been studied by
\cite{Braun1999} , who find that  this can have a significant effect on performance. 



In this paper, we are concerned with the task of monitoring for changes in the parameter
$\theta$ when there is no prior knowledge available regarding either its pre-
or post- change value. We use a test statistic which has an easily computable exact null distribution, which makes it suitable for
monitoring sequences which may not have a large number of available observations
between each change point. Further,,
our method is highly computationally efficient
and is hence suitable for deployment in situations where observations are being
received at high frequencies, such as the aforementioned data stream setting. We will focus on detecting an increase in the Bernoulli parameter since this tends to be more important in practice, although our methods are trivially extendable to two-sided change detection.

Our work is based on the Change Point Model (CPM) framework described in
\cite{Hawkins2003}, as a tool to extend traditional methods designed for fixed-length sequences to
 to the sequential
monitoring setting. The
CPM
was originally introduced to monitor for changes in a Gaussian mean, but
was
later extended to Gaussian variances \citep{Hawkins2005}, and nonparametric
distributional shifts \citep{Rosstechnometrics}. We propose to extend this work to monitor a
Bernoulli proportion.  We have made R code implementing our method publicly available: details on this can be found in the Software Implementation section at the end of this paper.


The remainder of the paper proceeds as follows. Section \ref{sec:cpm1} begins
with the problem of testing for a change in a Bernoulli parameter when dealing with a
fixed length sequence. Section \ref{sec:cpm2} then shows how the technique we develop
can be extended to sequential monitoring, where observations are being received over time. A
discussion of implementation issues follows in Section \ref{sec:computation},
and an empirical evaluation of performance is carried out in Section
\ref{sec:experiments}.

\section{The Bernoulli Change Point Model: Fixed Length Sequence}
\label{sec:cpm1}
We first consider the problem of detecting a change point in a fixed length
sequence. In this case there are $t$
Bernoulli observations,
$X_1,\ldots,X_t$. We stress again that neither the pre- or post- change value
of $\theta_t$ is assumed to be known, and we assume for now that the sequence
contains at most one change point.

For a specified point $k$ in this sequence, we can use a standard two-sample
hypothesis test to assess whether a change point occurs at $\tau=k$, with the null
hypothesis being that there is no change and that all $t$ observations are
identically distributed.

 Several such tests
exist in the statistics literature. We choose to use Fisher's Exact Test
\citep{Agresti1992}
(FET) since
its null distribution can be computed exactly, rather
than relying on
Gaussian approximations which only hold asymptotically. This
property is important since we would like our change detector to be deployable
in
situations where only a small number of observations are available between change points. Another
important property of FET is that its null distribution does not depend on
the true value of $\theta_t$.

The idea behind FET is as follows: suppose the observations at time $t$ are broken up into two samples $x_1,\ldots,x_k$ and $x_{k+1},\ldots,x_t$. Let the null hypothesis be that there are no
change points in the sequence, which implies that both samples have been generated by the same Bernoulli distribution with a fixed parameter $\theta_0$.
Under this assumption, the $X_i$ variables are
identically distributed, with $P(X_i = 1) =
\theta_0$,  and $P(X_i = 0) = 1 - \theta_0$ for all $i$. 
Let $S_t$ be a random variable defined as the number of failures in the first $t$ observations, i.e:

$$S_t = \sum_{i=1}^{t}X_i.$$
Then, conditional on $S_t = s_t$, FET uses a combinatorics argument to reason about how the number of observed failures are distributed between the two samples. Let $S_k$ be the number of failures in the first sample. Under the null hypothesis, the probability that $S_k=s_k$ follows a hypergeometric distribution:

\begin{equation}
P(S_k = s_k | S_t=s_t)=\frac{ {s_t \choose s_k} {t-s_t \choose k-s_k}}{{t\choose k}},
\label{eqn:hypergeometric}
\end{equation}
where $t \choose k$ is the binomial coefficient. A fundamental property of the FET is that this probability does not depend on the unknown parameter $\theta_0$. By conditioning on the value of the sufficient statistic $S_t$, this dependency has been removed. Therefore the p-values of the FET under the null hypothesis are independent of $\theta_0$, which makes this test suitable for situations where this parameter is not known. Now, as noted in the Introduction, we will generally be more interested in detecting an increase in $\theta_t$, which corresponds to an unusually small number of failures occuring within the first $k$ observations. The probability of there being $s_k$ or
less failures in the first $k$ observations under the null hypothesis that there is no change point and all observations are identically distributed is:

$$p_{k,t} = \sum_{i=1}^{k}P(S_k = s_i), \quad p_{k,t} \in [0,1].$$
This is the one-sided p-value of the FET. For convenience, we will instead work with the statistic $F_{k,t}$,
defined as:

$$F_{k,t} = 1-p_{k,t}, \quad F_{k,t} \in [0,1].$$
Finally, the null hypothesis that no change occurs at $k$ is rejected if
$F_{k,t} >
h_{k,t}$ for some appropriately chosen threshold  $h_{k,t}$.

Of course, since we have no prior knowledge of where the change point is
located, we do
not know which value of $k$ to use for testing. The null hypothesis is now that
 there is no change point in the data, while the alternative hypothesis is that
a change point exists at any location. To perform this
test,  we can use a method analogous to the procedure followed in generalized
likelihood ratio testing \citep{Pettitt1980}. We compute $F_{k,t}$ at every point
$1<k<t$,
and use the maximum
value. This leads to the statistic:

$$F_{t} = \max_{k} F_{k,t}, \quad 1 < k < t.$$
If  $F_{t} > h_t$ for some suitably chosen
threshold $h_t$, then the null hypothesis is rejected, and we conclude that
a change occured at some point in the data. In this case, the best
estimate
$\hat{\tau}$ of the location of the
change point is at the value of $k$ which maximized $F_{t}$. If $F_{t}
\leq h_t$, then we do not reject the null hypothesis, and hence conclude that no
change has occurred. The choice of this $h_t$
threshold will be discussed further in the following section.

\section{The Bernoulli Change Point Model: Sequential Monitoring}
\label{sec:cpm2}
We now consider the task of change detection when new
observations are being received in discrete time. Let $X_t$ be the $t^{th}$ observation
where $t \in {1,2,\ldots}$ is growing and perhaps unbounded. 

For each value of $t$, we can treat the sequence
$X_1,\ldots,X_t$ as being of a fixed length, compute $F_t$, and use the methodology from the previous section to test whether a change point has occurred. If a change is detected, we give a signal,and stop monitoring. If no change is detected, then we wait until the next observation $X_{t+1}$ is received and repeat this process, computing $F_{t+1}$ and
again testing for a change. In other words, we propose computing $F_{t}$ at each
time point, and signalling that a change has occurred when $F_{t} > h_t$. In the case where the sequence
may contain multiple change points, the monitoring process is then restarted
immediately at the observation following the change, with all previous observations being discarded. 

Although recomputing $F_t$ whenever a new observation is received may seem
 computationally expensive, it can actually be computed in a very efficient
manner,
as will be discussed in Section
\ref{sec:computation}.

The key issue with this approach then becomes determining the sequence of thresholds $\{h_t\}$. As mentioned in the Introduction, an important requirement for sequential monitoring algorithms is that the rate of false alarms can be bounded, where a false alarm constitutes flagging that a change has
occurred when no change has taken place. In other words, assuming that no change
has occurred, we
ideally would have:

\begin{equation}
P(F_{t} > h_t | F_{t-1} \leq h_{t-1},\ldots,F_{1} \leq h_1,\theta_t=\theta_0) =
\alpha,
\label{eqn:conditional}
\end{equation}
where $\alpha$ is some user-specified constant. In this case, the expected time
between false alarms (again denoted as the $ARL_0$, for Average Run Length)
is equal to $1 / \alpha$. Because the finite-sample conditional distribution of $F_t$ cannot generally be computed analytically, the usual approach when working with change point models is to use Monte Carlo simulation to compute the sequence of $h_t$ values corresponding to the desired $ARL_0$. This can be a computationally expensive procedure which is not feasible to carry out in real-time as data is being received, therefore the usual solution is to compute the $h_t$ values corresponding to many different choices of the $ARL_0$ in advance, and then store these in a lookup table so they can be easily used \citep{Hawkins2005, Rosstechnometrics}.

However, a problem arises --  unlike the examples considered in the existing CPM literature, the FET test involves conditioning on the observed number of successes, as can be seen from Equation \ref{eqn:hypergeometric}. This implies that the thresholds $h_t$ used in the CPM should be conditional on the particular data sequence $\{X_t\}$ that been observed, with different sequences requiring different thresholds. Since a collection of $n$ Bernoulli random variables has $2^n$ possible realisations, it is hence not possible to use the sort of precomputed lookup table which is used in \cite{Hawkins2005} and \cite{Rosstechnometrics}. 

As a solution to this problem we design the CPM in a conservative fashion. In general, $p_t$ is smallest (with $F_t$ being largest) when $\theta_0 = 0.5$.. Therefore, in order to generate a threshold sequence $h_t$ that will an give an $ARL_0$ of at least $1/\alpha$, we simulate Bernoulli sequences under the assumption that $\theta_0 = 0.5$ (full details of this simulation follow below). Because other values of $\theta_0$ result in lower values of $F_t$, this will result in an $ARL_0$ which is greater than $1/\alpha$, i.e. we will have the more conservative criteria:

\begin{equation}
P(F_{t} > h_t | F_{t-1} \leq h_{t-1},\ldots,F_{1} \leq h_1,\theta_t=\theta_0) \leq
\alpha.
\label{eqn:conditional2}
\end{equation}

One final complication which can arise is the discreteness of the test statistic $F_k$. Because the sequence of $X_t$ random variables has a Bernoulli distribution, there is only a finite number of values which $F_k$ can achieve. This can potentially add a further degree of conservativeness to the procedure.

In order to reduce the discreteness of the test statistic, we borrow an idea from \cite{Zhou2009} and recommend smoothing the $F_k$ values. Specifically, we define a new statistic $Y_t$ which is formed by applying exponential smoothing to the $F_{k,t}$ statistics, 

\begin{eqnarray}
 Y_{1,t}  &=& F_{1,t}    \nonumber \\
  Y_{k,t}  &=& (1-\lambda)Y_{k-1,t} + \lambda F_{k,t} \nonumber \\
 Y_{t}  &=&\max{k} Y_{k,t}, \quad \lambda \in [0,1].
\end{eqnarray}

The underlying idea is that if a change point occurs at $\tau$, then the
values of $F_{k,t}$ should be high whenever $k$ is close to $\tau$.
Therefore, smoothing the $F_{k,t}$ statistics in this way should not
negatively impact performance. Further, by using the smoothing, the range of
possible values for each $Y_{k,t}$ is significantly increased compared to the
$F_{k,t}$ statistics, hence $Y_{t}$ has more possible values than $F_t$. 
This allows a sequence of thresholds to be chosen which give an $ARL_0$
closer to the desired value. We are now faced with the issue of choosing $\lambda$; generally, a value close to $0$ results in more smoothing and produces a change detector which is slightly better at detecting smaller shifts in $\theta_t$, while a higher value results in more smoothing and is hence more suitable for detecting larger shifts. In our Experiments section we will explore several different choices of $\lambda$ and show that the performance difference is not overly large.

We now return to the original problem of finding a sequence of thresholds which satisfies Equation \ref{eqn:conditional2}. This is non-trivial; the marginal distribution of the $F_{t}$ and $Y_{t}$ statistics  is complex, and the
conditional distribution in Equation \ref{eqn:conditional} moreso. Since
it does not seem possible to determine the thresholds analytically, we
instead compute them using a Monte Carlo technique . In order to compute the thresholds for some choice of $\alpha$, we simulate one million streams each containing $2000$ Bernoulli($0.5$)  observations. For each stream, we compute $Y_{t}$ at each observation. The required values of $h_t$ can then be found successively, by starting with the first observation and choosing $h_1$ to make the proportion of $Y_1$ values exceeding $h_1$ equal to $\alpha$. The streams which exceed $\alpha$ are then discarded, and $h_2$ can then be chosen so that the proportion of remaining $Y_2$ values exceeding $h_2$ is equal to $\alpha$, and so on. In this way, the conditional distributions are successively approximated, allowing the threshold sequence to be found.

Although this simulation is computationally expensive and may require several hours of processing, this is not a problem since it only needs to be performed a single time, and can hence be performed in advance. Then, once we have computed these values, we can store them in a lookup table which can be accessed with no computational overhead. We present such a lookup table in Table \ref{tab:controllimits} in the Appendix, which gives the threshold sequences corresponding to several choices of $\alpha$. It can be seen that these
thresholds seem to slowly converge towards constant values,  so it seems reasonable to
use the value of $h_{2000}$ as an approximation of $h_t$ for $t>2000$.

\begin{table*}
\centering
\begin{tabular}{|c | c c| } 
\hline
$\theta_0$ & \multicolumn{2}{c|}{Empirical $ARL_0$}\\
\hline
& $\lambda=0.1$ & $\lambda=0.3$\\
\hline
0.01 & 971 (862)& 1039 (765)\\
0.05 & 622 (543)&638  (597)\\
0.10 & 589 (489)&529 (530)\\
0.20 & 512 (461)&516  (491)\\
0.30 & 509 (460)&509 (474)\\
0.40 & 504 (471)& 507 (483)\\
0.50 & 500 (493)& 500 (494)\\
\hline
\end{tabular}
\caption{Empirical $ARL_0$ when the CPM is designed with to have an $ARL_0$ of $500$, for several values of $\theta_0$. Standard deviations are shown in brackets.}
\label{tab:conservative}
\end{table*}

Since we have computed these thresholds under the assumption that $\theta_0=0.5$, the CPM will be conservative when the parameter of the Bernoulli sequence is not equal to this value, causing the $ARL_0$ to be higher than desired. In order to investigate just how conservative this procedure is, Table \ref{tab:conservative} shows the actual empirical $ARL_0$ values that are achieved by a CPM designed to have $ARL_0 = 500$, when $\theta_0$ takes on a variety of other values. It can be seen that the degree of conservativeness is quite small, unless $p_0$ takes on a very small value ($< $0.01). The CPM hence seems suitable for use on most Bernoulli streams, which we will investigate further in Section \ref{sec:experiments}.

\section{Implementation Issues}
\label{sec:computation}

Having completed the description of the Bernoulli CPM, we now turn towards its
implementation. In many important real world scenarios, computational resources are
limited so it is important to have a change detection algorithm which can be
computed efficiently.
For the CPM, the
majority of computation time is spent calculating the $F_{k,t}$ statistics.
From Equation \ref{eqn:hypergeometric}  we can see that this is equivalent to
evaluating the probability mass function of the hypergeometric distribution.
Most common statistical packages will provide a highly optimized routine for
this. However, we can increase computational efficiency by exploiting the high level of
correlation between the $F_{k,t}$ statistics.

Consider a fixed length sequence containing
$t$ observations, of which $s_t$ are failures. Let $s_k$ be the number of
failures observed before observation $k$. Write:
$$d_{k,t} =\frac{ {s_t \choose s_k} {t-s_t \choose k-s_k}}{{N
\choose n}}.$$

Now, rather than computing $d_{k+1,t}$ from scratch, we can compute it
recursively from $d_{k,t}$. We make use of the following identities for the
binomial
coefficient:

$${n \choose {k+1}} = {n \choose k}\frac{n-k}{k+1}, \quad {{n+1} \choose k} =
{n \choose k}\frac{n+1}{n+1-k}.$$
From this, algebraic manipulation shows that:

\begin{equation}
d_{k,t+1} = \left\{ \begin{array}{rl}
 \frac{d_{k,t} (s_t+1)(t-s_t+1)(t+1-s_t)}{(k+1-s_k)(t-s_t-k-s_k+1)(t+1)}
&\mbox{ if $X_{t+1}=1$} \\
  \frac{d_{k,t} (t-s_t+1)(t+1-s_t)}{(t-s_t-k-s_k+1)(t+1)}
&\mbox{ if $X_{t+1}=0$}
       \end{array} \right.
\label{eqn:rec1}
\end{equation}
And similarly: 

\begin{equation}
d_{k+1,t} = \left\{ \begin{array}{rl}
 \frac{d_{k,t} (s_t - s_k)(k+1)}{(s_k+1)(t-k)} &\mbox{ if $X_{k+1}=1$} \\
  \frac{d_{k,t} (t-s_t-k+s_k)(k+1)}{(k-s_k+1)(t-k)} &\mbox{ if $X_{k+1}=0$}
       \end{array} \right.
\label{eqn:rec2}
\end{equation}
Using these recursive formulations significantly decreases the
processing time required to compute each value of $F_{k,t}$.
Recall that $F_{k,t}$ is defined as:

$$F_{k,t} = 1- \sum_{i=1}^{k}d_{i,t}. $$
Therefore for all $k < t$, the value of $d_{k,t}$ can be calculated from
$d_{k,t-1}$
using Equation \ref{eqn:rec1}, without the need to evaluate any
factorials. When $i=t$ this does not apply, since $d_{t,t-1}$ is not defined. In
this case, $d_{k,t}$ can be computed from $d_{k-1,t}$ using Equation
\ref{eqn:rec2}. 

However even though this recursive formulation allows
efficient computation of
$F_{t}$, and hence the smoothed $Y_{t}$ statistics,
the time required to compute these values still grows linearly over time as
more observations are received. Further, it also requires all previous data
points to be retained in memory. In situations where this is not feasible due
to constraints on processing power/memory, we can use a further efficiency
device where only the previous $w$ observations are stored in memory, with the
previous $t-w$ discarded.

We define a \textbf{window} $W_{w,t}$ of length
$w$
to be the set of the $w$ most recently observed points, i.e.
$W_{w,t} = \{x_{t-w+1},\ldots,x_t\}$. Only the points in this window are stored
in
memory, with older points discarded.
The $F_{t}$ statistic is now calculated by maximising $F_{k,t}$ over only
the observations in the window, rather than over the whole stream. Therefore,
the memory only needs to
be large enough to contain the previous $w$ points, and the computational cost
of computing $F_{t}$ is constant rather than growing over time. 

A naive implementation of windowing where old points are discarded
would have a significant negative effect on the performance of the change detector.
Therefore, we do not discard old points entirely, but instead summarise them in
the sufficient statistic $s_{t-w}$ which maintains the sum of the observations too old to
be included in the window. 

In other words, suppose that at time $t$, there have been $s_{t-w}$ failures outside the current window. We define:

$$S_k = s_{t-w} + \sum^{k}_{i=t-w+1}X_i, \quad t-w < k < t .$$

The $F_{k,t}$ statistics are then defined for $t-w < k < t$, and maximisation is carried out only in this range. Because $s_{k,t}$ is a sufficient statistic for $X_1,\ldots,X_w$, no information is lost by discarding these points, and values of $F_t$ at each point in
the window is identical to the value when no windowing is used, meaning that no
information is lost, so no loss in performance will occur. Also because
older points are summarised in $s_{t-w}$, the choice of the window size $w$ is
not critical since it only determines the points at which a change may be
detected.

\section{Performance Analysis}
\label{sec:experiments}

Having introduced the Bernoulli CPM, we now proceed to evaluate its
performance. We compare it to the Bernoulli CUSUM chart introduced in
\cite{Reynolds1999}. Unlike the CPM, the CUSUM requires both
the pre- and post- change
values of $\theta_t$ to be known, and is the optimal change detector under this assumption, in the sense of minimising the worst case detection delay \citep{Lorden1971}. Since our
CPM assumes nothing about  $\theta_t$, we would expect its performance to be
inferior, so we consider the CUSUM as a benchmark which defines optimal
performance. Because designing an optimal CUSUM is unrealistic since generally the values of $\theta_0$ and $\theta_1$ will not be known, we also investigate the effect that misspecification of these parameters has on the CUSUM.

Following standard practice in the change detection literature
\citep{Basseville1993}, we evaluate
performance by setting the $ARL_0$ of both methods to be equal, and then compare
the
expected delay taken to detect changes of various magnitudes. In order to this we need to choose some value for the $ARL_0$, 
so we choose $ARL_0=500$ for both methods. Our results apply without loss of generality to other choices of the $ARL_0$.

Since our CPM assumes no prior knowledge of the pre-change parameter $\theta_0$ and
instead require it to be learned from the data, the
location of the change
point will affect their performance. A change which occurs early will be more
difficult to detect than one which occurs late, since fewer observations are
available to allow $\theta_0$ to be estimated accurately. We therefore consider
two
different change point locations, $\tau=\{50,300\}$, which correspond to
early and moderately located changes. In the case of sequences containing
multiple change points, this corresponds to how far apart the change points are
located.

We now briefly review the CUSUM before proceeding
with the performance comparison.

\subsection{CUSUM}
\label{sec:cusum}
Our implementation of the CUSUM chart follows \cite{Reynolds1999}. Given a
sequence
of Bernoulli observations $x_1,x_2,\ldots$ with known parameter $\theta_0$
before the change point and $\theta_1$ after, we define:

$$C_0 = 0$$
$$C_t = \max(0,C_{t-1} + x_t - k),$$
where $k$ is a reference value defined as $k = r_1/r_2$, where:

$$r_1 = -\log\left(\frac{1-\theta_1}{1-\theta_0}\right),\quad r_2 =
\log\left(\frac{\theta_1(1-\theta_0)}{\theta_0(1-\theta_1)}\right).$$
A change is flagged when $C_t > h(\theta_0,\theta_1)$ for some appropriately
chosen control limit $h(\theta_0,\theta_1)$, which is chosen in order to give the CUSUM a specified $ARL_0$. This can be done using (for example) the approximation scheme discussed in \cite{Reynolds1999}, or by Monte-Carlo simulation. We note that due to the discreteness of the test statistic, it will generally not be possible to achieve a target $ARL_0$ exactly. Although we previously discussed this problem in the context of our CPM, it is more serious for the CUSUM since the $C_t$ statistic is more discrete. For each choice of $\theta_0$ and $\theta_1$, we computed control limits which gave the CUSUM the minimum possible $ARL_0$ greater than $500$. In practice however, the achieved $ARL_0$s were in the range $500$-$600$, but for cases where the test statistic is especially discrete, most notably when $\theta_0=0.1$ and $\theta_1=0.9$, it is not possible to find any threshold greater between $400$ and $800$.

\subsection{Results and Discussion}

We wish to compare the expected delay taken by both the CPM and the CUSUM to detect changes of various magnitues in a Bernoulli parameter. We consider cases where the pre-change value of the Bernoulli parameter $\theta_0$ takes values from the set $\{0.1,0.2,0.3,0.4\}$, and the post-change value is $\theta_1 \in \{\theta_0+0.1,\theta_0 + 0.2,\ldots,0.9\}$. By the symmetry of the Bernoulli distribution, the results for $\theta_0 = \gamma$ will be identical to those for $\theta_0 = 1-\gamma$, so we do not consider cases where $\theta_0 > 0.5$.

\begin{table*}[t]
\centering
\footnotesize{
\begin{tabular}{|c | c | c c c c c c c c| } 
\hline
& &\multicolumn{8}{c}{$\theta_1$}\\ 
\hline
Detector & $\theta_0$ & $0.2$ & $0.3$ & $0.4$ & $0.5$ & $0.6$ & $0.7$ & $0.8$ & $0.9$ \\
\hline
\multirow{4}{*}{CUSUM}
&  0.1 &   43.9 &   19.5 &   12.0 &    8.0 &    6.1 &    5.0 &    4.1 &    3.5 \\
&  0.2 &   0.00 &   58.8 &   25.7 &   15.5 &   10.3 &    7.4 &    5.7 &    4.5 \\
&  0.3 &   0.00 &   0.00 &   66.7 &   29.1 &   16.8 &   12.0 &    8.7 &    6.0 \\
&  0.4&   0.00 &   0.00 &   0.00 &   70.8 &   31.0 &   17.6 &   11.4 &    8.4 \\
\hline

\multirow{4}{*}{FET CPM, $\lambda=0.1$}
&  0.1 &   55.6 &   21.4 &   13.2 &    9.5 &    7.5 &    6.1 &    5.2 &    4.5 \\
&  0.2 &   0.00 &   77.7 &   28.2 &   16.5 &   11.7 &    9.0 &    7.4 &    6.2 \\
&  0.3 &   0.00 &   0.00 &   90.0 &   32.0 &   18.5 &   13.0 &    9.9 &    8.1 \\
&  0.4 &   0.00 &   0.00 &   0.00 &   98.1 &   34.1 &   19.4 &   13.4 &   13.1 \\

\hline
\multirow{4}{*}{FET CPM, $\lambda=0.3$}
&  0.1 &   57.4 &   21.0 &   12.3 &    8.6 &    6.6 &    5.3 &    4.5 &    3.9 \\
&  0.2 &   0.00 &   81.0 &   28.1 &   15.7 &   10.7 &    8.0 &    6.4 &    5.3 \\
&  0.3 &   0.00 &   0.00 &   98.0 &   32.4 &   17.7 &   11.8 &    8.7 &    6.9 \\
&  0.4 &   0.00 &   0.00 &   0.00 &  106.9 &   34.7 &   18.5 &   12.1 &    9.0 \\

\hline

\end{tabular}

}
\caption{Expected delay taken to detect a change from $\theta_0$
to $\theta_1$ when the change occurs at point $\tau=300$. Standard deviations are provided in the Appendix.}
\label{tab:chap3_bernoulli300results}
\end{table*}

For each choice of the parameters $\theta_0$ and $\theta_1$, and the change time $\tau$, we generated $20000$ sequences of Bernoulli($\theta_t$) observations where $\theta_t = \theta_0$ when $t < \tau$, and $\theta_t = \theta_1$ when $t \geq \tau$. We then configured the CUSUM to be optimal for these parameter values, by choosing $k$ as described in Section \ref{sec:cusum}. The CPM of course does not have any knowledge of these parameter values. For each sequence, we then ran both the CPM and the CUSUM, and found the time $T$ at which a change was flagged. The expected delay is then $E[T | T > \tau]$. Recall that the CPM uses a parameter $\lambda$ which controls the degree of smoothing applied to the $F_{k,t}$ statistics; we ran the CPM using both $\lambda=0.1$ and $\lambda=0.3$ to investigate the effect this has on performance.

\begin{table*}[t]
\centering
\footnotesize{
\begin{tabular}{|c | c | c c c c c c c c| } 
\hline
& &\multicolumn{8}{c}{$\theta_1$}\\ 
\hline
Detector & $\theta_0$ & $0.2$ & $0.3$ & $0.4$ & $0.5$ & $0.6$ & $0.7$ & $0.8$ & $0.9$ \\
\hline
\multirow{4}{*}{CUSUM}
&  0.1 &   43.7 &   19.4 &   12.0 &    8.0 &    6.1 &    5.0 &    4.1 &    3.5 \\
&  0.2 &   0.00 &   58.2 &   25.5 &   15.5 &   10.2 &    7.4 &    5.7 &    4.5 \\
&  0.3 &   0.00 &   0.00 &   66.1 &   29.4 &   16.8 &   12.0 &    8.7 &    6.0 \\
&  0.4&   0.00 &   0.00 &   0.00 &   70.5 &   30.8 &   17.6 &   1145 &    8.5 \\
\hline

\multirow{4}{*}{FET CPM, $\lambda=0.1$}
&  0.1 &  163.2 &   41.8 &   17.8 &   11.3 &    8.5 &    6.8 &    5.7 &    4.9 \\
&  0.2 &   0.00 &  201.5 &   63.0 &   23.8 &   13.9 &   10.0 &    7.9 &    6.6 \\
&  0.3 &   0.00 &   0.00 &  224.3 &   75.5 &   27.9 &   15.5 &   10.9 &    8.4 \\
&  0.4 &   0.00 &   0.00 &   0.00 &  238.8 &   84.0 &   29.3 &   15.7 &   11.0 \\
\hline
\multirow{4}{*}{FET CPM, $\lambda=0.3$}
&  0.1 &  169.8 &   43.5 &   17.0 &   10.2 &    7.4 &    5.8 &    4.8 &    4.1 \\
&  0.2 &   0.00 &  210.1 &   68.1 &   24.0 &   13.0 &    9.0 &    6.8 &    5.5 \\
&  0.3 &   0.00 &   0.00 &  237.0 &   85.3 &   28.4 &   14.6 &    9.6 &    7.3 \\
&  0.4 &   0.00 &   0.00 &   0.00 &  252.8 &   96.8 &   30.7 &   14.8 &    9.8 \\
\hline

\end{tabular}

}
\caption{Expected delay taken to detect a change from $\theta_0$
to $\theta_1$ when the change occurs at point $\tau=50$. Standard deviations are provided in the Appendix.}
\label{tab:chap3_bernoulli50results}
\end{table*}

The expected delays are presented in Table \ref{tab:chap3_bernoulli300results} for the case where $\tau=300$, and Table \ref{tab:chap3_bernoulli50results} for $\tau=50$. Several aspects of these results deserve comment:

\begin{itemize}
 \item As expected, the CPM performs substantially better when the change occurs after $300$ observations, than when it occurs after only $50$. This will be the case with any change detection algorithm which does not have full knowledge of the pre-change distribution and must instead learn it online. When only $50$ observations are available, it is difficult to form an accurate estimate of $\theta_0$ and performance suffers. 
\item The CPM does not seem overly sensitive to the choice of $\lambda$ used for smoothing. When $\lambda=0.1$, the CPM is slightly better at detecting small changes, while being worse at detecting larger ones, since the increased smoothing causes the large jump in the test statistic to be partially flattened out. As in the case of EWMA charts, the `best' choice of $\lambda$ will depend on the magnitude of change for which it is most important to minimize the detection delay.

\item The performance of the CPM is very close to the CUSUM when the change occurs after $300$ observations, demonstrating its usefulness. However, it is markedly worse when the change occurs after only $50$ observations. It is important not to read too much into the relatively poor performance of the CPM in the latter case, since the CUSUM is of course designed under full knowledge of the Bernoulli parameters. It is hence only a measure of the best possible performance, when everything is known exactly. We will explore this further in the following section.
\end{itemize}

\subsection{Effect of Parameter Misspecification on the CUSUM}

The CUSUM used above was designed based on complete knowledge of the true values of $\theta_0$ and $\theta_1$, and under these assumptions it is the optimal sequential change detector, in the minimax sense described by \cite{Lorden1971}. However in practice this optimality is of dubious value, since it is rare that these parameters will be known exactly, and practical deployments of the CUSUM must take into account the possibility of parameter misspecification. 

\begin{table*}[t]
\centering
\footnotesize{
\begin{tabular}{|c | c | c c c c c c c| } 
\hline
& &\multicolumn{7}{c}{$\theta_1$}\\ 
\hline
Detector & $\theta_0$ & $0.2$ & $0.3$ & $0.4$ & $0.5$ & $0.6$ & $0.7$ & $0.8$ \\
\hline
\multirow{4}{*}{Misspecified CUSUM}
 & 0.1 &  497.0 &   70.5 &   30.1 &   16.6 &   11.0 &    8.0 &    6.0 \\
 & 0.2 &   0.00 &  484.3 &   80.2 &   33.0 &   20.0 &   13.2 &    8.7 \\
 & 0.3 &   0.00 &   0.00 &  487.2 &   87.2 &   35.5 &   19.7 &   13.3 \\
 & 0.4 &   0.00 &   0.00 &   0.00 &  482.4 &   89.0 &   36.3 &   19.8 \\

\hline

\end{tabular}

}
\caption{Average delay required to detect a change from $\theta_0$
to $\theta_1$ when the change occurs at point $\tau=50$. Standard deviations are provided in the Appendix.}
\label{tab:cusummisspec}
\end{table*}

 In order to get a more realistic picture of how the CPM compares to the CUSUM, we now investigate the effect that such parameter misspecification has on the performance of the CUSUM. This is most likely to happen when the change occurs early during the monitoring, since there may be insufficient observations to allow $\theta_0$ to be accurately estimated. We therefore consider the case where the change occurs at observation $\tau=50$. Suppose that instead of being designed based on the true value of $\theta_0$, the CUSUM is instead designed based on $\tilde{\theta_0} = \theta_0 + 0.1$, which corresponds to a small degree of misspecification of the pre-change parameter. In this case, the CUSUM will be using a slightly higher value for the control limit $h$ than is optimal, and there should hence be a decrease in performance. Similarly, we assume that the post-change value of the parameter is also misspecified, and that the CUSUM is designed under the assumption that $\tilde{\theta_1} = \theta_1 + 0.1$. So for example, in a situation where the parameter changes from a value of $0.3$ to $0.6$, the CUSUM is designed under the false assumption that it is changing from $0.4$ to $0.7$. This is a relatively small degree of misspecification which could easily occur in practice.

The expected delay was computed in the same way as in the previous section, and the results are shown in Table \ref{tab:cusummisspec}. Comparing this to Table \ref{tab:chap3_bernoulli50results} in the previous section shows that the performance of the misspecified CUSUM is substantially worse than the CPM across every value of $\theta_0$ and $\theta_1$. The CPM therefore seems to be a more appropriate change detector in situations where $\theta_0$ may not be estimated accurately, and where the exact magnitude of the change is unknown.

\section{Conclusions}
We developed a change point model for Bernoulli sequences where new
observations are being received over time, and the pre-change value of the
parameter is unknown. Several computational devices were introduced which
allows the relevant test statistics to be computed recursively in a very
efficient manner. Our method has very favorable performance compared to
the optimal CUSUM chart. When the change occurs relatively late in the stream, the CPM gives comparable performance to the CUSUM chart which had full knowledge of the pre- and post- change Bernoulli parameters. But when the change occurs early in the stream, the CPM method performs poorly compared to CUSUM. However, in many data stream settings there will not be full knowledge of these parameters, and hence this optimal CUSUM is not very realistic. When we investigated how the CUSUM performs under small degrees of parameter misspecification, it's performance is seen to decline drastically, and the CPM seems the best tool in this unknown parameter situation.

\section{Software Implementation}

An implementation of the CPM methodology described in this paper can be found in the \textbf{cpm} R package, which is available either from CRAM (http://cran.r-project.org), or from the first author's website (http://gordonjross.co.uk).

\section{Acknowledgements}

This research was undertaken as part of the ALADDIN (Autonomous Learning Agents
for Decentralised Data and Information Systems) project and is jointly funded
by a BAE Systems and EPSRC (Engineering and Physical Research Council)
strategic partnership, under EPSRC grant EP/C548051/1. 

\bibliographystyle{asa}
\bibliography{jabref}

\pagebreak
\appendix
\section{Additional Tables}

\begin{table}[ht]
\centering
\footnotesize{

\begin{tabular}{| c |  c  c  c c |c c c c|}
\hline
&\multicolumn{8}{c|}{$ARL_0$}\\ 
\hline
&\multicolumn{4}{c}{$\lambda=0.1$} & \multicolumn{4}{|c|}{$\lambda=0.3$}\\ 
\hline

t & 370 & 500 & 1000 & 5000 & 370 & 500 & 1000 & 5000 \\
\hline
20 & 0.9232 & 0.9284 & 0.9474 & 0.9620 & 0.9700 & 0.9735 & 0.9801 & 0.9867\\
21 & 0.9144 & 0.9247 & 0.9318 & 0.9524 & 0.9657 & 0.9703 & 0.9774 & 0.9872\\
22 & 0.9091 & 0.9138 & 0.9321 & 0.9531 & 0.9627 & 0.9684 & 0.9767 & 0.9870\\
23 & 0.9048 & 0.9156 & 0.9254 & 0.9500 & 0.9626 & 0.9672 & 0.9766 & 0.9888\\
24 & 0.8999 & 0.9109 & 0.9249 & 0.9501 & 0.9622 & 0.9679 & 0.9769 & 0.9892\\
25 & 0.9009 & 0.9071 & 0.9273 & 0.9500 & 0.9631 & 0.9686 & 0.9783 & 0.9890\\
26 & 0.8971 & 0.9087 & 0.9247 & 0.9517 & 0.9640 & 0.9695 & 0.9792 & 0.9902\\
27 & 0.8974 & 0.9066 & 0.9250 & 0.9523 & 0.9642 & 0.9702 & 0.9797 & 0.9911\\
28 & 0.8964 & 0.9051 & 0.9259 & 0.9522 & 0.9645 & 0.9706 & 0.9809 & 0.9912\\
29 & 0.8958 & 0.9071 & 0.9260 & 0.9538 & 0.9650 & 0.9706 & 0.9812 & 0.9920\\
30 & 0.8966 & 0.9057 & 0.9268 & 0.9549 & 0.9658 & 0.9718 & 0.9817 & 0.9931\\
40 & 0.9057 & 0.9179 & 0.9392 & 0.9643 & 0.9712 & 0.9771 & 0.9857 & 0.9956\\
50 & 0.9199 & 0.9317 & 0.9509 & 0.9742 & 0.9759 & 0.9809 & 0.9886 & 0.9966\\
60 & 0.9303 & 0.9411 & 0.9597 & 0.9817 & 0.9777 & 0.9826 & 0.9904 & 0.9976\\
70 & 0.9381 & 0.9489 & 0.9657 & 0.9859 & 0.9794 & 0.9842 & 0.9918 & 0.9979\\
80 & 0.9430 & 0.9536 & 0.9698 & 0.9888 & 0.9807 & 0.9854 & 0.9923 & 0.9983\\
90 & 0.9470 & 0.9575 & 0.9738 & 0.9904 & 0.9812 & 0.9860 & 0.9929 & 0.9984\\
100 & 0.9486 & 0.9591 & 0.9758 & 0.9918 & 0.9821 & 0.9867 & 0.9934 & 0.9985\\
200 & 0.9599 & 0.9696 & 0.9840 & 0.9962 & 0.9844 & 0.9892 & 0.9945 & 0.9990\\
300 & 0.9631 & 0.9728 & 0.9860 & 0.9971 & 0.9848 & 0.9891 & 0.9950 & 0.9992\\
400 & 0.9637 & 0.9731 & 0.9868 & 0.9974 & 0.9852 & 0.9888 & 0.9952 & 0.9992\\
500 & 0.9652 & 0.9735 & 0.9876 & 0.9976 & 0.9854 & 0.9897 & 0.9953 & 0.9992\\
600 & 0.9654 & 0.9743 & 0.9873 & 0.9977 & 0.9847 & 0.9889 & 0.9954 & 0.9994\\
700 & 0.9639 & 0.9747 & 0.9876 & 0.9978 & 0.9856 & 0.9896 & 0.9954 & 0.9993\\
800 & 0.9668 & 0.9757 & 0.9881 & 0.9979 & 0.9858 & 0.9896 & 0.9953 & 0.9993\\
900 & 0.9669 & 0.9761 & 0.9885 & 0.9981 & 0.9859 & 0.9897 & 0.9953 & 0.9993\\
1000 & 0.9671 & 0.9763 & 0.9811 & 0.9982 & 0.9860 & 0.9897 & 0.9954 & 0.9994\\
2000 & 0.9679 & 0.9767 & 0.9892 & 0.9984 & 0.9861 & 0.9899 & 0.9955 & 0.9994\\

\hline
\end{tabular}
}
\caption{Value of the threshold $h_t$ achieving various values of the
$ARL_0$ for the Bernoulli CPM}
\label{tab:controllimits}
\end{table}

\begin{table*}[t]
\centering
\footnotesize{
\begin{tabular}{|c | c | c c c c c c c c| } 
\hline
& &\multicolumn{8}{c}{$\theta_1$}\\ 
\hline
Detector & $\theta_0$ & $0.2$ & $0.3$ & $0.4$ & $0.5$ & $0.6$ & $0.7$ & $0.8$ & $0.9$ \\
\hline
\multirow{4}{*}{CUSUM}
&  0.1 &   35.2 &   14.2 &    8.0 &    5.1 &    3.5 &    2.6 &    1.8 &    1.2 \\
&  0.2 &   0.00 &   47.6 &   18.8 &   10.1 &    6.1 &    4.3 &    2.8 &    1.6 \\
&  0.3 &   0.00 &   0.00 &   52.6 &   19.6 &   10.5 &    6.8 &    4.1 &    2.5 \\
&  0.4 &   0.00 &   0.00 &   0.00 &   56.1 &   21.6 &   10.7 &    6.2 &    3.6 \\

\hline

\multirow{4}{*}{FET CPM, $\lambda=0.1$}
&  0.1 &   53.7 &   15.3 &    8.1 &    5.1 &    3.6 &    2.6 &    2.0 &    1.5 \\
&  0.2 &   0.00 &   80.4 &   20.5 &    9.7 &    6.0 &    4.2 &    3.0 &    2.2 \\
&  0.3 &   0.00 &   0.00 &  101.6 &   23.4 &   10.5 &    6.4 &    4.3 &    2.9 \\
&  0.4 &   0.00 &   0.00 &   0.00 &   99.7 &   25.7 &   10.9 &    6.2 &    4.0 \\

\hline
\multirow{4}{*}{FET CPM, $\lambda=0.3$}
& 0.1 &   64.9 &   16.1 &    8.2 &    5.0 &    3.4 &    2.3 &    1.7 &    1.2 \\
& 0.2 &   0.00 &   89.6 &   22.5 &   10.1 &    5.9 &    3.8 &    2.5 &    1.8 \\
& 0.3 &   0.00 &   0.00 &  115.2 &   25.9 &   11.0 &    6.2 &    3.8 &    2.3 \\
& 0.4 &   0.00 &   0.00 &   0.00 &  113.2 &   28.5 &   11.6 &    5.9 &    3.3 \\

\hline

\end{tabular}

}
\caption{Standard deviations accompanying Table 1, for changes from $\theta_0$ to $\theta_1$ occurring at location $\tau=300$.}
\label{tab:chap3_bernoulli300resultssd}
\end{table*}

\begin{table*}[t]
\centering
\footnotesize{
\begin{tabular}{|c | c | c c c c c c c c| } 
\hline
& &\multicolumn{8}{c}{$\theta_1$}\\ 
\hline
Detector & $\theta_0$ & $0.2$ & $0.3$ & $0.4$ & $0.5$ & $0.6$ & $0.7$ & $0.8$ & $0.9$ \\
\hline
\multirow{4}{*}{CUSUM}

&  0.1 &   35.2 &   14.2 &    8.0 &    5.1 &    3.5 &    2.6 &    1.8 &    1.2 \\
&  0.2 &   0.00 &   47.6 &   18.8 &   10.1 &    6.1 &    4.3 &    2.8 &    1.6 \\
&  0.3 &   0.00 &   0.00 &   52.6 &   19.6 &   10.5 &    6.8 &    4.1 &    2.5 \\
&  0.4 &   0.00 &   0.00 &   0.00 &   56.1 &   21.6 &   10.7 &    6.2 &    3.6 \\
\hline
\multirow{4}{*}{FET CPM, $\lambda=0.1$}
&  0.1 &  186.0 &   70.1 &   19.3 &    8.1 &    4.9 &    3.4 &    2.5 &    1.9 \\
&  0.2 &   0.00 &  204.2 &  101.3 &   30.2 &   10.3 &    5.9 &    3.8 &    2.7 \\
&  0.3 &   0.00 &   0.00 &  212.2 &  114.5 &   38.3 &   12.2 &    6.1 &    3.8 \\
&  0.4 &   0.00 &   0.00 &   0.00 &  217.1 &  122.2 &   37.2 &   11.0 &    5.4 \\

\hline
\multirow{4}{*}{FET CPM, $\lambda=0.3$}
&  0.1 &  192.2 &   73.5 &   21.6 &    7.8 &    4.6 &    3.0 &    2.1 &    1.4 \\
&  0.2 &   0.00 &  212.4 &  111.5 &   30.6 &   10.5 &    5.7 &    3.2 &    2.1 \\
&  0.3 &   0.00 &   0.00 &  216.5 &  127.4 &   42.5 &   12.3 &    5.6 &    3.1 \\
&  0.4 &   0.00 &   0.00 &   0.00 &  227.6 &  136.2 &   45.8 &   11.0 &    4.7 \\

\hline

\end{tabular}

}
\caption{Standard deviations accompanying Table 2, for changes from $\theta_0$ to $\theta_1$ occurring at location $\tau=50$.}
\label{tab:chap3_bernoulli50resultssd}
\end{table*}

\begin{table*}[t]
\centering
\footnotesize{
\begin{tabular}{|c | c | c c c c c c c| } 
\hline
& &\multicolumn{7}{c}{$\theta_1$}\\ 
\hline
Detector & $\theta_0$ & $0.2$ & $0.3$ & $0.4$ & $0.5$ & $0.6$ & $0.7$ & $0.8$ \\
\hline
\multirow{4}{*}{Misspecified CUSUM}
& 0.1 &  245.9 &   58.9 &   22.8 &   11.3 &    7.0 &    4.7 &    2.8 \\
& 0.2 &   0.00 &  244.2 &   66.6 &   24.6 &   13.6 &    7.9 &    4.8 \\
&  0.3 &   0.00 &   0.00 &  240.5 &   73.5 &   27.0 &   13.0 &    7.7  \\
&  0.4 &   0.00 &   0.00 &   0.00 &  242.1 &   74.2 &   26.8 &   13.0  \\

\hline

\end{tabular}

}
\caption{Standard deviations accompanying Table 3, for changes from $\theta_0$ to $\theta_1$ occurring at location $\tau=50$.}
\label{tab:cusummisspecsd}
\end{table*}

\end{document}